\documentclass[aps,prl,reprint,onecolumn,12pt,groupedaddress]{revtex4-1}

\usepackage[utf8]{inputenc} 
\usepackage[T1]{fontenc}
\usepackage{graphicx} 
\usepackage{amsmath }
\usepackage{booktabs} 
\usepackage{array} 
\usepackage{paralist}  
\usepackage{verbatim}  
\usepackage{lmodern}
\usepackage{braket}
\usepackage{float}
\newcommand{\pf}{0.65}

\begin{document}

\title{Phase super-resolution with N00N states\\ generated by on demand single-photon sources}
\author{M.\,Müller}
\email[Electronic adress:\ ]{m.mueller@ihfg.uni-stuttgart.de}
\author{H.\,Vural}
\author{P.\,Michler}
\affiliation{Institut f\"ur Halbleiteroptik und Funktionelle Grenzfl\"achen, Center for Integrated Quantum Science and Technology (IQ$^\textrm{ST}$) and SCoPE, University of Stuttgart,, Allmandring 3, 70569 Stuttgart, Germany.}
\date{\today}
\pagestyle{empty}
\thispagestyle{empty}

\begin{abstract}
Multi-photon entangled states such as `N00N states' have attracted a lot of attention because of their possible application in high-precision phase measurements. 
So far, N00N states have been generated by spontaneous parametric down-conversion sources and by mixing quantum and classical light on a beam splitter. 
However, single-photon on demand sources promise to be more efficient and scalable than down-conversion sources which are probabilistic in nature. 
Here, we demonstrate super-resolving phase measurements based on two-photon N00N states generated by a quantum dot single-photon source utilizing the Hong-Ou-Mandel effect on a beam splitter. 
The quantum dot was excited through a two-photon resonant excitation scheme. 
Using a $\pi$-pulse, this results in the deterministic generation of indistinguishable single photons from both, the biexciton and exciton radiative recombination processes. 
Launching either type of the photons into both inputs of an interferometer, phase super-resolution, i.e. an interference fringe spacing with doubled rate relative to the single photon rate has been observed for both transitions. 
Interference visibilities of $V_{N=2,\mathit{XX}}=0.54\pm0.01$ for the biexciton and $V_{N=2,\mathit{X}}=0.46\pm0.02$ for the exciton transition have been obtained, respectively. 
\end{abstract}

\maketitle

Optical quantum metrology provides a route to enhance sensing applications by utilizing e.g. non-classical states of light\,\cite{Giovannetti2006,Dowling2008,Taylor2016}. 
For many photonic sensing applications, a general task is measuring a phase $\varphi$ with a precision $\Delta \varphi$. 
Entangled photon states promise to beat the shot noise limit or so-called standard quantum limit (SQL) which can be reached for example using a laser source in an interferometric sensing scheme. 
The SQL relates the error of phase estimation $\Delta \varphi$ with the photon number $N$ used for the measurement to $\Delta \varphi = 1/\sqrt{N}$. 
A maximally path-entangled multi-photon state, a so-called N00N state $\ket{\psi}_{N}=\frac{1}{\sqrt{2}}(\ket{N,0}+\ket{0,N})$, acquires a phase at a rate $N$ times as fast as classical light, referred to as super-resolution\,\cite{Jacobsen1995,Fonseca1999}. 
Employing these states, high precision optical phase measurements can be performed achieving the Heisenberg limit $\Delta \varphi = 1/N$ which outperforms the SQL by $1/\sqrt{N}$, often referred to as super-sensitivity\,\cite{Caves1981,Kuzmich1998}.  

So far, various schemes for generating N00N states have been realized and phase super-resolution has been demonstrated in a number of studies\,\cite{Rarity1990, Bouwmeester2004, Walther2004, Mitchell2004}. 
Phase super-sensitivity, or beating the SQL has been demonstrated with four-entangled photons using state projection to study the N00N component of various initial $N$-photon states\,\cite{Nagata2007}. 
The largest N00N state generated to date contained 5 photons by mixing quantum and classical light\,\cite{Afek2010}. 
Up to now, all schemes have used spontaneous parametric down-conversion sources which are inherent probabilistic with Poissonian photon statistics. 
In contrast, on-demand solid state single-photon emitters are better suited for the generation of entangled photon states due to their higher efficiency and brightness\,\cite{Dousse2010,Claudon2010}. In addition, compact on-chip implementation is also feasible\,\cite{Arcari2014,Rengstl2015}.

\begin{figure}[H]
\begin{center} 
\includegraphics[width=\pf\linewidth]{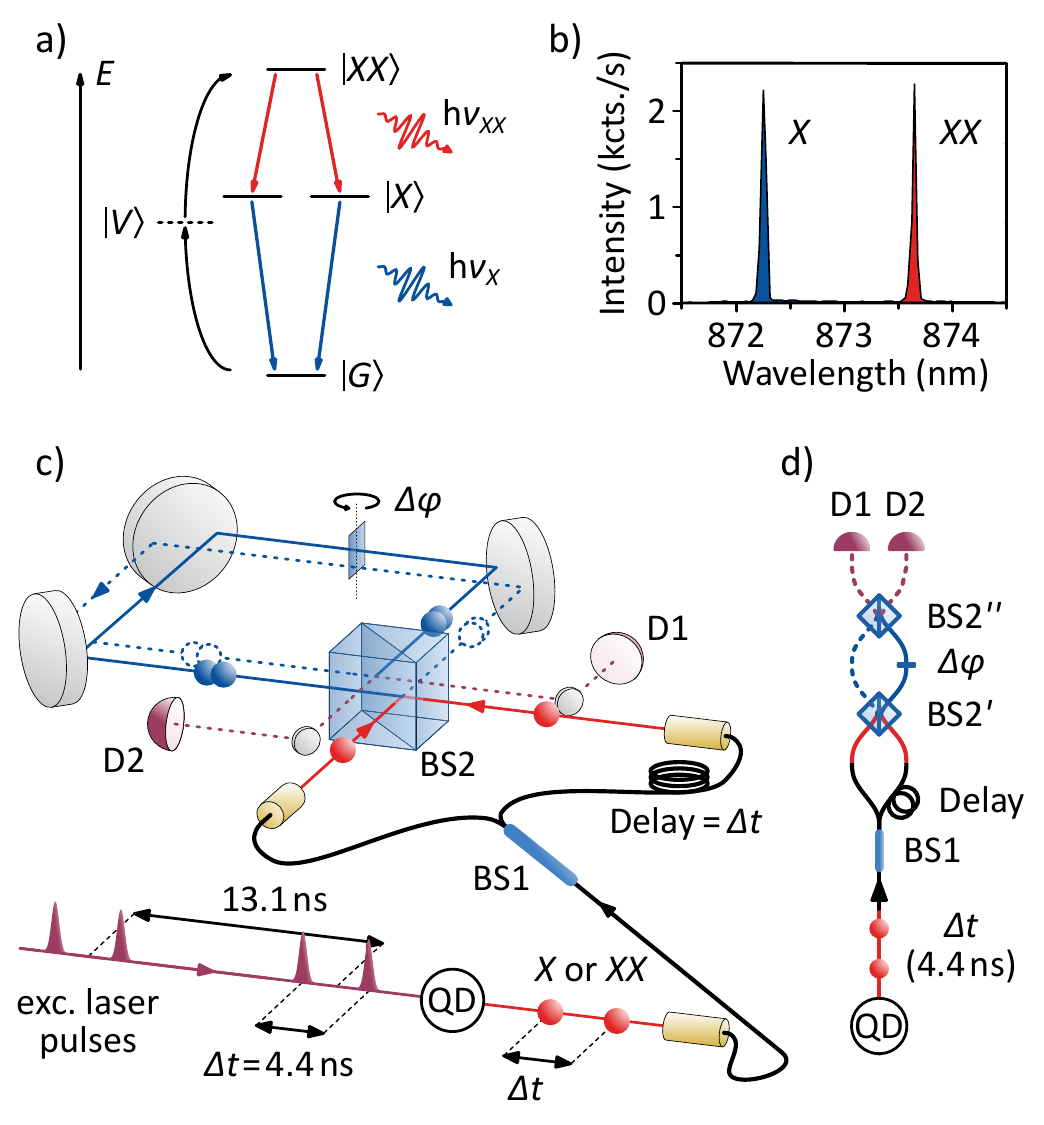} 
\caption{
a) Two-photon excitation scheme via the virtual state $\ket{V}$ and the subsequent cascaded emission of a pair of biexcitonic ($\textrm{h}\nu_{\textit{XX}}$) and excitonic ($\textrm{h}\nu_{X}$) single-photons. 
b) Emission spectrum in resonant two-photon excitation. 
The pulsed laser, energetically sitting exactly in between the \textit{X} and \textit{XX} emission line, is completely suppressed (see Methods).
c) Intrinsically phase-stable double-path Sagnac interferometer\,\cite{Nagata2007}, together with the laser excitation scheme.
Initially BS2 serves, together with BS1, as the second part of an unbalanced Mach-Zehnder interferometer to generate the $\frac{1}{\sqrt{2}}(\ket{20}+\ket{02})$ state. 
The path-entanglement is then probed via phase dependent autocorrelation measurements again on BS2, on a second interference position BS2$''$. 
The relative phase difference $\Delta \varphi$ between the two paths is varied by rotating a phase plate. 
d) Deconvoluted scheme of the photon path in the Sagnac interferometer. 
\label{fig1}} 
\end{center} 
\end{figure}

We present an experimental realization of the simultaneous generation of two two-photon N00N states with a single quantum emitter by using the radiative biexciton (\textit{XX}) -- exciton (\textit{X}) cascade in a single semiconductor quantum dot (QD). 
The QD is excited by a resonant two-photon absorption process\,\cite{Brunner1994,Stufler2006,Jayakumar2013,Muller2014} ensuring a deterministic and coherent excitation (see Fig.\,\ref{fig1}\,a),b) and Supplementary Information). 
The subsequently emitted photons possess perfect single-photon emission statistics and an adequate indistinguishability ($V_{\textrm{HOM},\textit{XX}}=0.76\pm0.03$ and $V_{\textrm{HOM},X}=0.50\pm0.04$) which allows the production of the two two-photon N00N states and for the observation of phase super-resolution.

\begin{figure}[H]
\begin{center} 
\includegraphics[width=\pf\linewidth]{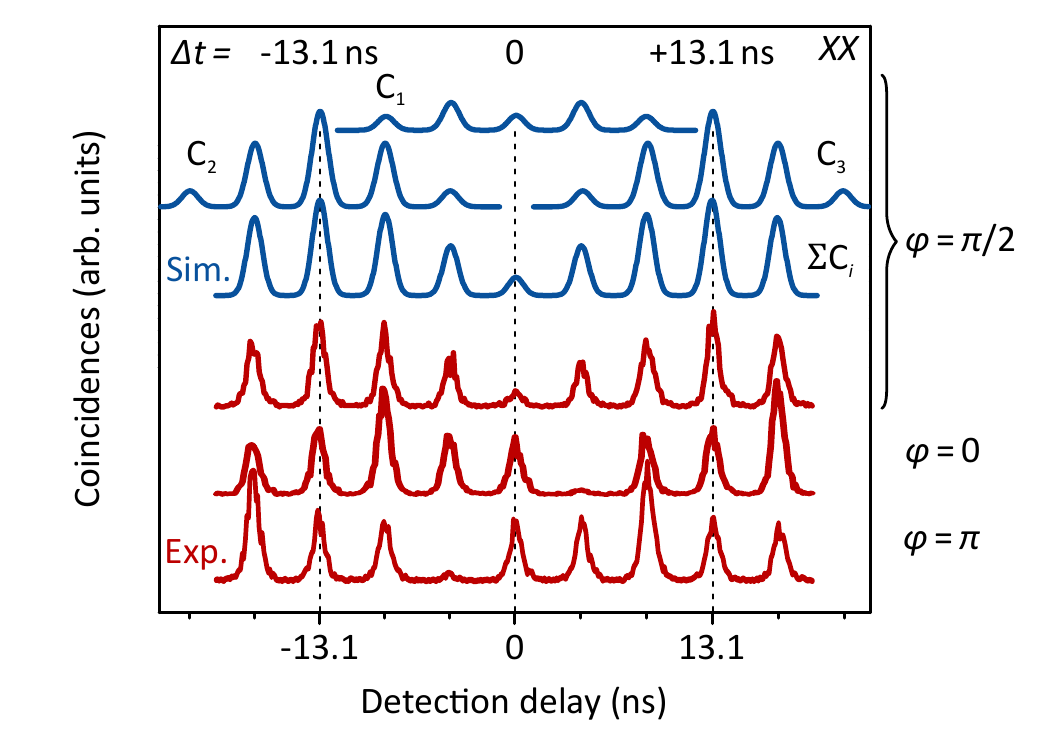} 
\caption{
Coincidence histogram simulation and data for the \textit{XX} measurement. The blue lines C1, C2, and C3 show the expected five peak clusters which are separated by the laser repetition period of 13.1\,ns. For a phase shift of $\varphi=\pi/2$ the distribution of the coincidence probabilities correspond to the scenario of a single beam splitter and hence to a typical Hong-Ou-Mandel experiment. The sum of these clusters gives the expected histogram which is in very good agreement with the measurement (top red line). The two red lines for phase shifts of $\varphi=0$ and $\varphi=\pi$ reflect measurement situations of constructive biphotonic interference.
\label{fig2}} 
\end{center} 
\end{figure}

\subparagraph{Results}
The experimental setup is illustrated in Fig.\,\ref{fig1}\,c),d).
By a specific measurement design (see Method section) two successive emitted photons from either the \textit{X} or \textit{XX} decay are launched simultaneously onto BS2 from different input ports (red spheres), traveling through the Sagnac interferometer and are finally detected on detector D1 or D2.
For the coincidence histogram (see Fig.\,\ref{fig2}) we expect clusters of five peaks separated by the pump laser repetition period (13.1\,ns). 
The two photons arrive at BS2 with delays of $-2\Delta t, -\Delta t, 0, \Delta t$ and $2\Delta t$ with $\Delta t=4.4$\,ns.
Because of the 13.1\,ns delay between the clusters, the two outer peaks of the cluster temporally overlap with the the corresponding peaks from the previous/successive cluster.
However, the central peak, which reflects the desired situation when both photons arrive at the same time at beam splitter BS2 is well resolved and not affected by any overlap to the $\Delta t$ time separation between the pulses. 
It exhibits a minimum for $\varphi=\pi/2$ (top red line) and maxima for $\varphi=0, \pi$ (bottom red curves), expressing destructive and constructive biphotonic interference.
The complete phase dependent behavior is discussed below (further details about the occurrence and oscillating behavior of the coincidence peaks at non-zero detection delay as well as the theoretical analysis can be found in the Supplementary Material). 
At BS2$'$ the Hong-Ou-Mandel (HOM) effect\,\cite{Hong1987} for identical bosonic particles causes the generation of the two-particle path-entangled state
\begin{equation*}
\ket{\psi}_2=\frac{1}{\sqrt{2}}\left(\ket{2}_{c}\otimes\ket{0}_{d}+\ket{0}_{c}\otimes\ket{2}_{d}\right)=\frac{1}{\sqrt{2}}\left(\ket{20}+\ket{02}\right)\,,
\label{}
\end{equation*}
which implies that the photons can only be detected together, and also solely in either one of the two exit ports $c$ or $d$ of BS2$'$ (blue spheres).
The biphotonic N00N state evolves by passing through the Sagnac type double-path interferometer in which a relative phase $\textrm{e}^{\textrm{i}\varphi}$ is acquired in one mode, introduced by turning an ordinary glass plate. 
Because of the non-classical nature, the photonic state pics up the phase $N=2$ times faster than a coherent state would do\,\cite{Gerry2005}:
\begin{equation*}
\begin{aligned}
\ket{\psi}_2\stackrel{\varphi}{\rightarrow}\frac{1}{\sqrt{2}}\left(\ket{20}+\textrm{e}^{\textrm{i}2\varphi}\ket{02}\right)\,.
\end{aligned}
\label{eq:psi_2}
\end{equation*}  
The coherence of this state can then be determined by measuring the coincidence probability on detector D1 and D2 after BS2$''$:
\begin{equation}
\mathcal{P}_{\textrm{D1},\textrm{D2}}=\frac{1}{2}(1+\cos(2\varphi))\,.
\label{eq:N_2N_2}
\end{equation}
Thus, an oscillating behavior according to twice of the imprinted phase is expected in the autocorrelation measurement.

\begin{figure}[H]
\begin{center} 
\includegraphics[width=\pf\linewidth]{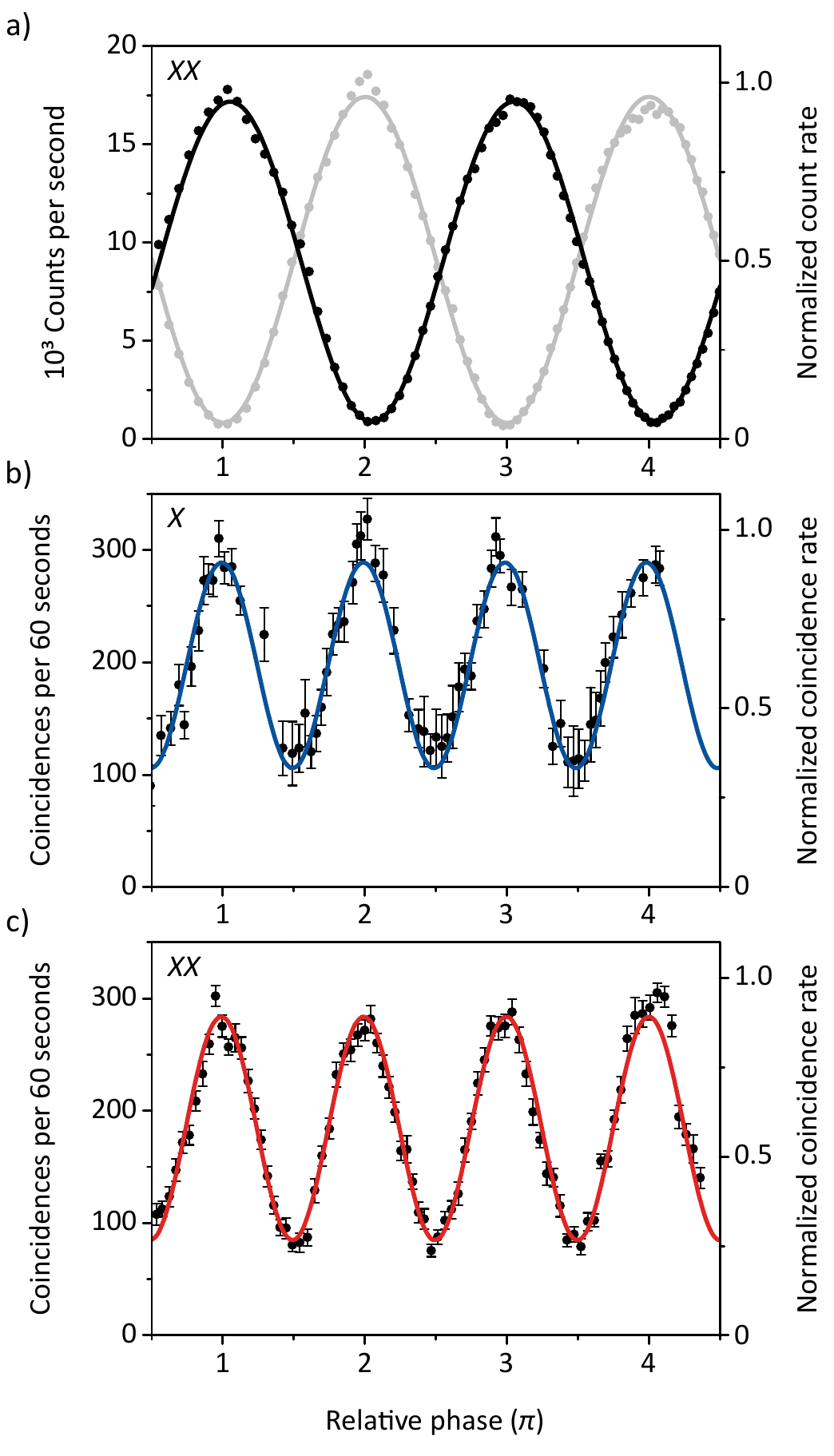} 
\caption{
a) Phase dependent \textit{XX} single-photon count rate for the detectors D1 (black) and D2 (gray). The signal is oscillating according to the photon frequency $\nu_{\textit{XX}}$.
b) Phase dependent \textit{X} coincidence rate. A doubling of the frequency is observed compared to the single-photon case. 
c) Phase dependent \textit{XX} coincidence rate. As in the previous graph, phase super-resolution is established with an even higher contrast.
\label{fig3}} 
\end{center} 
\end{figure}

\begin{figure}[H]
\begin{center} 
\includegraphics[width=\pf\linewidth]{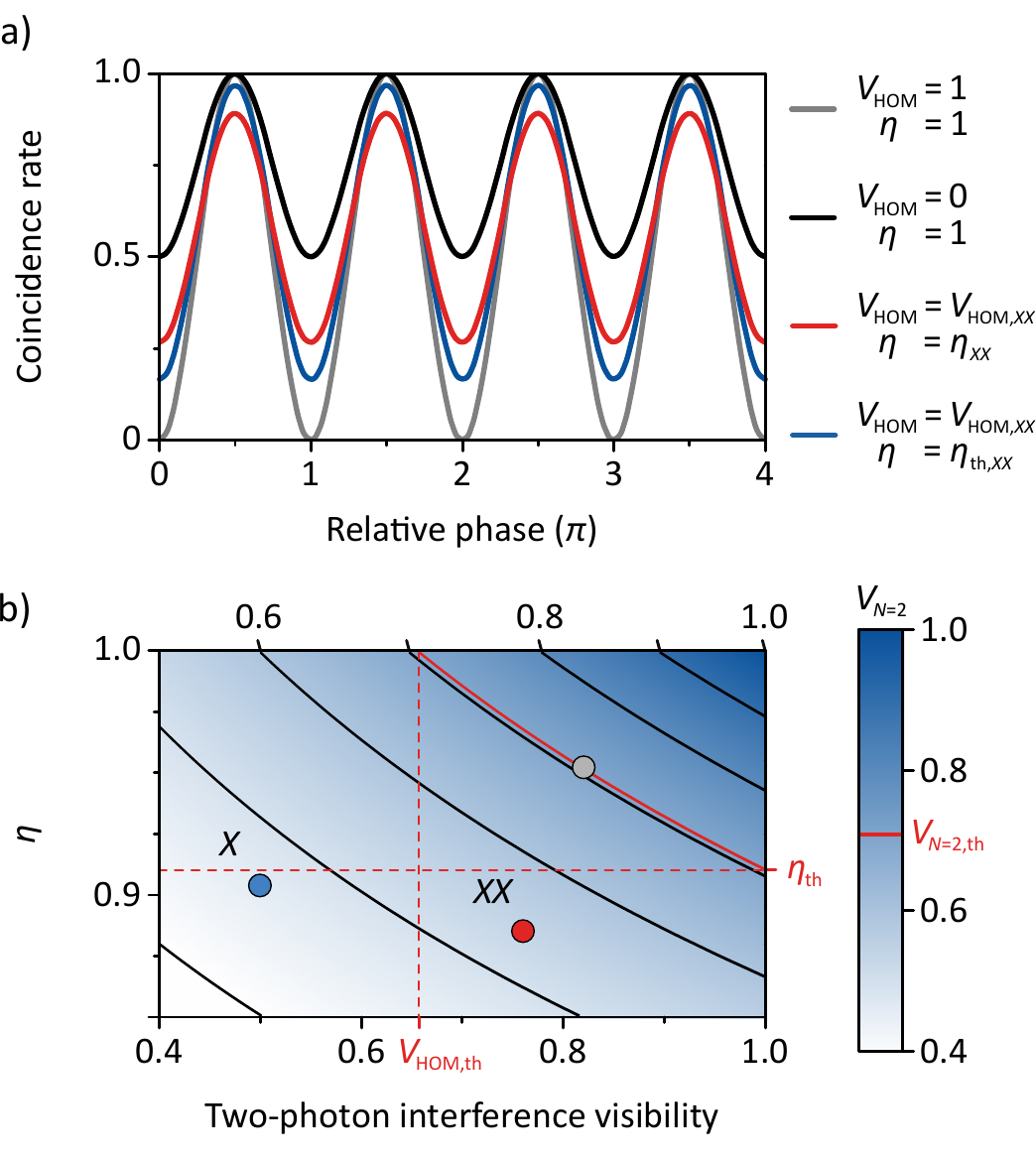} 
\caption{
a) Simulated phase dependent coincidence rate of a biphotonic N00N state for different degrees of indistinguishability of the individual photons $V_{\textrm{HOM}}$ and different Mach-Zehnder mode overlaps $\eta$. The red solid line corresponds to the measured \textit{XX} parameters, the blue one indicates the boundary to beat the SQL.
b) N00N state interference visibility as a function of $V_{\textrm{HOM}}$ and $\eta$. The blue and the red dot represent the measurements of \textit{X} and \textit{XX}. For the highest reported value of two-photon interference visibility from remote QDs\,\cite{Gao2013}, a Mach-Zehnder mode overlap of 0.95 would be sufficient to beat the SQL (gray dot).
\label{fig4}} 
\end{center} 
\end{figure}

Considering the actually relevant phase dependency of the correlations at zero detection delay, Fig.\,\ref{fig3} gives a deeper insight into the interference properties.
Fig.\,\ref{fig3}\,a) displays the intensity signal of a single \textit{XX} photon input (achieved by blocking one input of BS2$'$) for the two different detectors D1 and D2 as a function of the imprinted phase.
As expected, the respective intensities oscillate opposite in phase with a frequency of $\nu_{\mathit{XX}}$ and show a high contrast of $V_{N=1,\textit{XX}}=0.91\pm0.01$, only limited by the imperfect mode overlap on the beam splitter.
Comparing these single-photon detections with the coincidence signals for excitonic (Fig.\,\ref{fig3}\,b)) and biexcitonic (Fig.\,\ref{fig3}\,c)) biphotonic states, the reduced de-Broglie wavelength of $\lambda_{h\nu}/2$ is evident.
The pronounced oscillations show remarkably high coincidence rates of up to 300 coincidence per minute and reveal visibilities of $V_{N=2,\textit{XX}}=0.54$ and $V_{N=2,X}=0.46$ (see also Table\,\ref{tab1}).
It is striking, that both data sets do not oscillate around a mean value of $1/2$.
To obtain more insight on the underlying physical effects, the common beam splitter and phase transformations were applied to theoretically reproduced the phase dependent coincidence oscillations (solid lines). 
With this, the data can be modeled by the following expression (see Supplementary Information)
\begin{equation*}
\mathcal{P}^{\textrm{exp}}_{\textrm{D1},\textrm{D2}}=\frac{1}{4}\left(2+\eta^2(1-\eta^2V_{\textrm{HOM}})+\eta^2(1+\eta^2V_{\textrm{HOM}})\cos(2\varphi)\right)\,,
\label{eq:N_2N_2_exp}
\end{equation*}
which reduces to equation\,(\ref{eq:N_2N_2}) for the ideal case of $V_{\textrm{HOM}}=\eta=1$.
Here, $\eta$ combines the spatial mode overlap on BS2$'$ and BS2$''$ which form together the Mach-Zehnder measuring device\,\cite{Santori2002,Kacprowicz2010}.
To understand the implications of this relation, it is helpful to consider the following cases (see Fig.\,\ref{fig4}\,a)).
Assuming optimal experimental conditions $\eta=1$, perfectly indistinguishable photons $V_{\textrm{HOM}}=1$ would give rise to oscillations with unity contrast.
On the contrary entirely distinguishable photons ($V_{\textrm{HOM}}=0$) would lead to oscillations of half the amplitude around a mean value of $3/4$, confirming that frequency doubling can be achieved without any path-entangled state\,\cite{Nagata2007}, however limited to a maximum visibility of $1/3$.
The photon indistinguishability given by the two-photon output probability on BS2$'$ (typically called M-Value in literature), can be directly determined from the minima of the oscillations, and essentially defines the visibility of the N00N state interference.  
Only when taking into account the above introduced experimental imperfections $\eta$, results exhibiting maximum values below unity (especially the data for \textit{XX}) are explicable.

\subparagraph{Discussion}
These results clearly state, that the generation of biphotonic N00N states with a reduced de-Broglie wavelength is possible by utilizing a semiconductor QD as a deterministic source of identical single-photons.
It should, however, be noted that although phase super-resolution is definitely attested, the criterion for the so called phase super-sensitivity\,\cite{Nagata2007,Giovannetti2004} is not fulfilled, yet.
To be exact, in the applied creation process in which maximally $1/4$ of the photons can be utilized for the N00N state generation (in $3/4$ of the cases, the photons do not impinge simultaneously on BS2$'$), the threshold interference visibility of $V_{N=2,\textrm{th}}=1/\sqrt{N}=1/\sqrt{2}$ can not be beaten in principal.
This is a very specific constraint on the scheme utilized in this work and not a fundamental limitation of the source itself.
Employing for instance photons from remote QDs to the input ports of BS2 could overcome this problem.
The feasibility and demands of such an experiment are indicated in Fig.\,\ref{fig4}\,b).
Here the N00N state visibility $V_{N=2}$ is plotted as a function of the two-photon interference visibility $V_{\textrm{HOM}}$ and the Mach-Zehnder mode overlap $\eta$.
The respective threshold values are indicated with red lines.
Apart from the before mentioned principal constraints to beat the SQL, the \textit{X} measurement is limited by both, the photon indistinguishability as well as the experimental imperfections.
On the other hand, the result for the \textit{XX} photons excels the threshold value $V_{\textrm{HOM,th}}$, whereas a minimum of $\eta_{\textrm{th},\mathit{XX}}=0.97$ would be required to pass the boundary of the SQL.
Such values are typical for a mode overlap on a single beam splitter and can in principal also be achieved for a Mach-Zehnder interferometer by using for example single-mode fibers in the detection.  
Assuming the up to now highest reported indistinguishability visibility ($V_{\textrm{HOM}}=0.82$) from remote QD sources published in ref\,\cite{Gao2013}, a value of $\eta=0.95$ would be sufficient to truly beat the SQL.

Regarding the improvement in phase resolution by comparing the absolute wavelength of pump and signal of the source, the scheme used in this work gains a factor of two ($\lambda_{\textrm{pump}}\rightarrow\lambda_{\textrm{N00N}}=\lambda_{\textrm{pump}}/2$).
A spontaneous parametric down-conversion source instead will attain no improvement at all by generating a $N$\,=\,2 N00N state ($\lambda_{\textrm{pump}}\rightarrow2\cdot\lambda_{\textrm{signal/idler}}\rightarrow\lambda_{\textrm{N00N}}=\lambda_{\textrm{pump}}$). 
To achieve the same phase resolution improvement factor obtained by the QD single-photon source, in a down-conversion scheme a `higher' N00N state with a dimension of $N$\,=\,4 has to be generated\,\cite{Bouwmeester2004}.
It is worth to note, that through the \textit{XX}-\textit{X}-cascade, two photons are available per excitation cycle.
This could be exploited to increase the effective detection rate (and hence shorten the measurement time), or just by utilizing one of the photons as a heralding signal allowing for diverse measurement schemes.

In summary, we have demonstrated phase super-resolution of biphotonic N00N states by utilizing photons emitted from the \textit{X} and \textit{XX} state of a semiconductor quantum dot.
In a theoretical analysis we could reproduce the data and show the potential and limitations of such a device concerning optical phase measurements with a precision beating the SQL.
If recent results of transform-limited\,\cite{Kuhlmann2015} and near-unity indistinguishable\,\cite{He2013,Ding2016,Somaschi2015,Unsleber2015} photons from resonantly excited QDs can be successfully transferred to systems of remote emitters, it would facilitate transgressing this boundary in future experiments.    

\subparagraph{Methods}
The experiment presented in this study involved two main methodical elements. 
That is, the preparation of the quantum emitter and thus the generation of pure single and indistinguishable photons, as well as the creation and interferometric examination of path-entangled superposition states. 
\paragraph{QD preparation}
The InGaAs QD was optically excited via a two-photon process. 
In this way the biexciton state is deterministically populated and decays subsequently via cascaded emission of a biexcitonic and excitonic single-photon to the ground state. 
The excitation laser was operated in pulsed mode, providing 3\,ps pulses with a repetition rate of 76\,MHz. 
These pulses were temporally prolonged to 25\,ps in a self-built pulse stretcher. 
Although the laser energy is spectrally separated from the QD emission, residual scattered light on the detector is inevitable. 
However, techniques like an orthogonal excitation-detection geometry, cross-polarization suppression and utilization of interferential notch filters allow for an entire laser background free observation of the QD emission. 
For an efficient incoupling of the laser light from the side into the planar waveguide structure (one pair of GaAs/AlAs layers as distributed Bragg reflector on top and 15 pairs on the bottom side of a planar $\lambda$-cavity) a high NA-objective $(\textrm{NA}=0.55)$ was chosen. 
The emitted light was collected with a long working distance objective $(\textrm{WD}=17 \textrm{mm},\textrm{NA}=0.45)$ from the top and sent through a single-mode fiber to the analyzing part of the setup. 
\paragraph{N00N state generation and verification}
The creation of a path-entangled N00N state with $N$\,=\,2 using photons emitted from a QD single-photon source, is based on the method of ref\,\cite{Santori2002}. 
The excitation laser pulses are split up in a free-space Mach-Zehnder interferometer (MZI) with a variable path-length difference of the two arms, resulting in double pulses, temporally separated by the adjustable parameter $\Delta t$. 
Exciting the QD deterministically, two subsequent single-photon pairs from the \textit{XX}-\textit{X}-cascade are emitted with the same time separation $\Delta t$. 
After spectral filtering of one of the transitions, the photons were sent into the Sagnac interferometer (see Fig.\ref{fig1}\,c),d)). 
The fiber-coupled beam splitter BS1 and beam splitter BS2$'$ form together a MZI with a fixed path-length difference of $\Delta t_{\textrm{MZI}}=4.4$\,ns. 
The variable time delay $\Delta t$ is adjusted to match this fixed delay, in order to enable the probability that the subsequent photons take different paths in the MZI and impinge on BS2$'$ from different sides at the same time. 
If the photons are identical, the probability amplitudes of finding the photons in different output modes of the beam splitter interfere destructively. 
The photons will take always the same exit port and the wave function of the evolved state can therefore be described by the desired biphotonic N00N state.
Traveling through the Sagnac interferometer, the phase $\varphi$ between the two paths can be set to a precise value.
The coherence of the N00N state is then probed via coincidence measurements after BS2$''$ on the detectors D1 and D2 for different phase settings.

\vspace{0.5mm}
The authors thank L. Wang, A. Rastelli and O. G. Schmidt for providing the high-quality sample and acknowledge financial support from the center for Integrated Quantum Science and Technology (IQST) and the Deutsche Forschungsgemeinschaft (DFG) via the project MI 500/23-1.

\begin{table}[H]
\begin{center}
\setlength{\tabcolsep}{4mm}
\begin{tabular}{c*{6}{c}}
\toprule
						& $g^{(2)}(0)$ 	& $\eta$ 				& $\eta_{\textrm{th}}$& $V_{\textrm{HOM}}$ 	& $V_{N=1}$		 	& $V_{N=2}$ 		 \\
\midrule
\textit{X}	& $0.01\pm0.01$  & $0.90\pm0.01$ & -- 									& $0.50\pm0.04$ 			& $0.93\pm0.01$ & $0.46\pm0.02$ \\
\textit{XX} & $0.01\pm0.01$ & $0.89\pm0.01$ & $0.97$							& $0.76\pm0.03$ 			& $0.91\pm0.01$ & $0.54\pm0.01$ \\
\bottomrule
\end{tabular}
\caption{
Summarized measurement results for \textit{X} and \textit{XX}, respectively. $g^{(2)}(0)$ gives the multi-photon emission probability, $\eta$ describes the spatial beam overlap of the whole Mach-Zehnder interferometer, $\eta_{\textrm{th}}$ the theoretical beam overlap to beat the SQL, $V_{\textrm{HOM}}$ the two-photon interference visibility and $V_{N=i}$ the N00N state interference visibility for a $i=\{1,2\}$ state.}
\label{tab1}
\end{center} 
\end{table}

\end{document}


\title{Supplementary Information to \\`Phase super-resolution with N00N-states \\ generated by on demand single-photon sources'}
\author{M.\,Müller}
\email[Electronic adress:\ ]{m.mueller@ihfg.uni-stuttgart.de}
\author{H.\,Vural}
\author{P.\,Michler}
\affiliation{Institut f\"ur Halbleiteroptik und Funktionelle Grenzfl\"achen, Center for Integrated Quantum Science and Technology (IQ$^\textrm{ST}$) and SCoPE, University of Stuttgart,, Allmandring 3, 70569 Stuttgart, Germany.}
\date{\today}
\maketitle
\pagestyle{empty}
\thispagestyle{empty}

\begin{figure}[H]
\begin{center} 
\includegraphics[width=\linewidth]{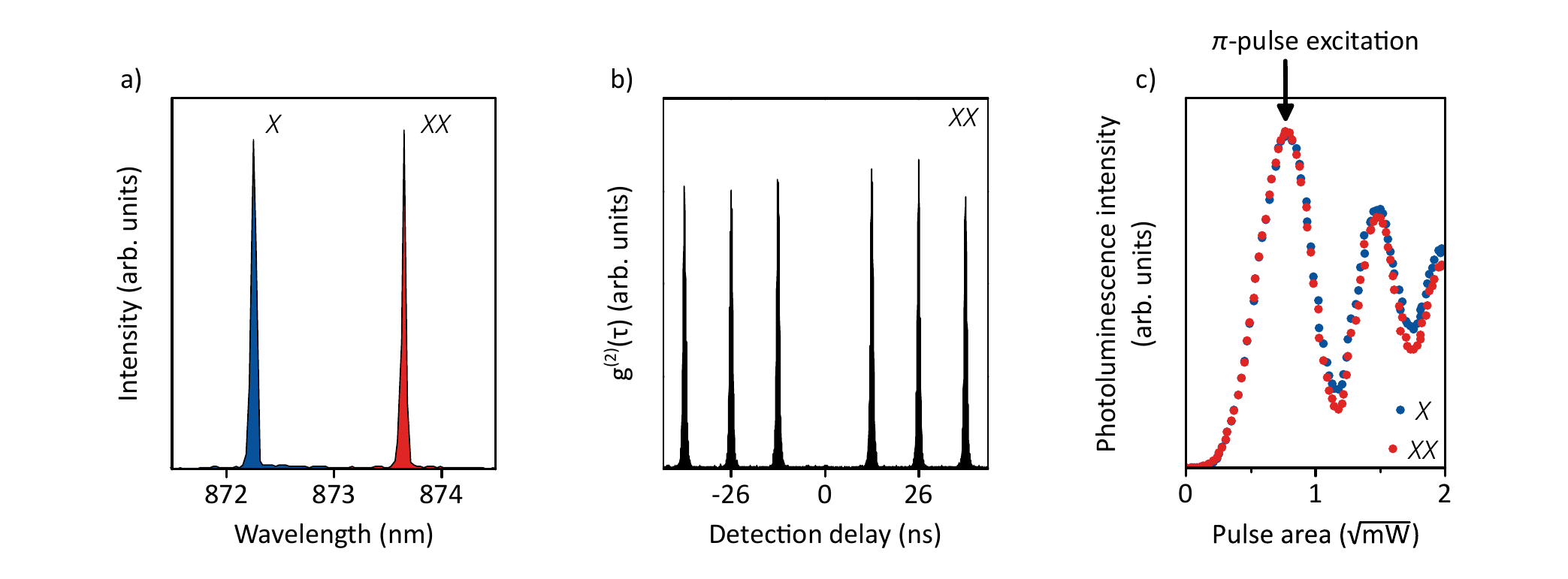} 
\caption{
a) Emission spectrum in resonant two-photon excitation. The pulsed laser, energetically sitting exactly in between \textit{X} and \textit{XX}, is completely suppressed. b) Second-order autocorrelation of the biexcitonic photons showing pure single-photon emission. The same holds for the excitonic photons (not shown here). c) \textit{X} and \textit{XX} state preparation. Highly pronounced Rabi rotations demonstrate the coherence of the two-photon excitation process. All measurements are carried out at the $\pi$-pulse, the maximum of the state preparation probability.  
\label{smfig1}} 
\end{center} 
\end{figure}

\subparagraph{Deterministic single-photon pair generation}
Fig.\,\ref{smfig1} illustrates the main characteristics of the deterministically operated single-photon pair source. 
Due to a different Coulomb interaction between the individual particles of the respective states, the energies of the emitted photons differ by the \textit{XX} binding energy ($2.3$\,meV), as shown in the emission spectrum in Fig.\,\ref{smfig1}a). 
This allows for two-photon optical excitation\,\cite{Brunner1994,Stufler2006,Jayakumar2013,Muller2014}, in which the laser energy is set precisely in between the quantum dot (QD) spectral lines to hit the two-photon resonance of the biexciton state. 
Various methods were applied to fully remove the scattered laser light from the detector (see Methods section of the main paper). 
As expected from the cascaded decay process, both spectral lines have almost equal intensity and moreover no other emission from the sample was detected across a wide spectral range. 
The very selective resonant excitation process facilitates extremely pure single-photon generation by the individual transitions, as evident from the second-order autocorrelation measurement of the biexcitonic photons depicted in Fig.\,\ref{smfig1}b). 
Similar degree of purity was observed for the excitonic photons (not shown here). 
Additionally, the resonantly driven ground state - biexciton system experiences Rabi rotations of the state occupation probabilities, monitored by the photoluminescence intensity, as a function of the excitation strength (see Fig.\,\ref{smfig1}c)). 
In a pulsed measurement, this quantity is given by the pulse area which in turn is identified with the square root of the incident excitation power. 
Apart from the highly pronounced oscillation, the measurement clearly shows a maximum of state preparation, the so called $\pi$-pulse, which was the working point for all measurements presented in the this letter. 
Under these conditions, the two-photon driven QD acts as deterministically single-photon source of high purity.  

\begin{figure}[H]
\begin{center} 
\includegraphics[width=\linewidth]{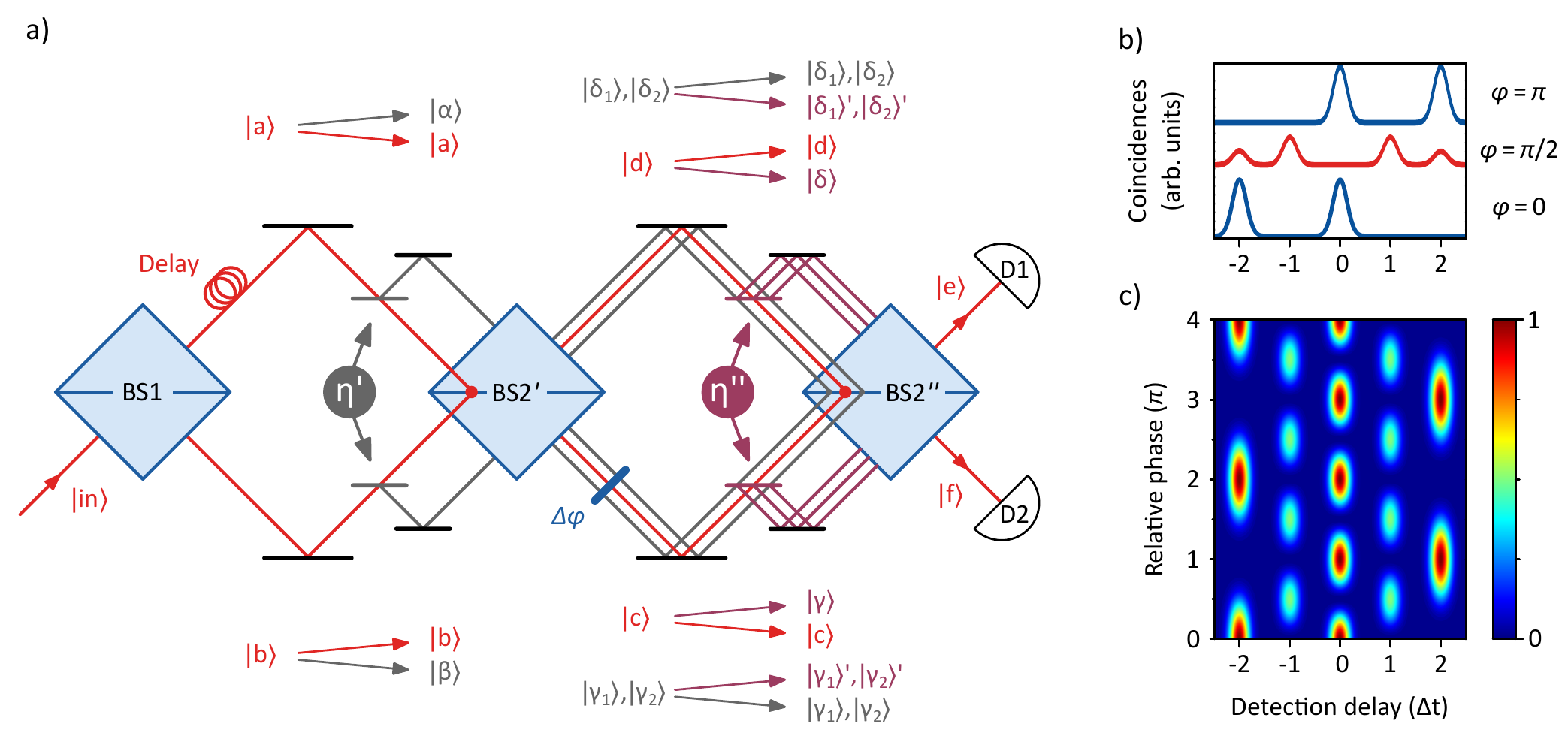} 
\caption{
a) Schematic of the experimental setup, including all possible spatial modes. $\eta'$ and $\eta''$ correct for non-perfect overlap on BS2$'$ and BS2$''$, respectively. 
The output probabilities are calculated for every possible coincidence combination.
b) Simulated histograms for selected phase settings, ignoring coincidences from consecutive excitation cycles. The well known Hong-Ou-Mandel type five peak pattern is found for integral multiples of $\varphi=\pi/2$. 
Here, entirely indistinguishable photons are assumed, leading to a complete suppression of the middle peak.
c) Simulated phase dependent histograms in a pseudocolor plot. 
The x-axis is given in units of the time difference between two pulses $\Delta t$.
Distinct oscillations with single ($\Delta t=\pm2$) and doubled ($\Delta t=0, \pm1$) frequencies are observable.
Though, biphotonic N00N-states only contribute to peaks for $\Delta t=0$, revealing phase super-resolution associated with a true reduction of the de-Broglie wavelength.
\label{smfig2}} 
\end{center} 
\end{figure}

\subparagraph{Theoretical description}
The theoretical modeling of the Mach-Zehnder operation for different state inputs was derived by combining the basic transformations of the beam splitter operator $\hat{B}$ (with $T=R=0.5$), phase operator $\hat{\varphi}$ and the operator $\hat{\eta}$ for beam overlap consideration (see below):
\begin{equation*}
\hat{B}=\frac{1}{\sqrt{2}}
\begin{pmatrix}
1 & \textrm{i} \\
\textrm{i} & 1
\end{pmatrix}, 
\quad \hat{\varphi}=
\begin{pmatrix}
\textrm{e}^{\textrm{i} \varphi} & 0  \\
0 & 1
\end{pmatrix}, 
\quad \hat{\eta}=
\begin{pmatrix}
\sqrt{\eta} & \textrm{i}\sqrt{1-\eta}  \\
\textrm{i}\sqrt{1-\eta} & \sqrt{\eta}
\end{pmatrix}
\label{seq1}.
\end{equation*}
Fig.\,\ref{smfig2} shows a sketch of the complete deconvoluted pathway of the photons with the respective modes. 
In the ideal case of perfectly indistinguishable photons ($V_\textrm{HOM}=1$) and maximum beam overlap ($\eta=1$), the transformation for the creation operators $\hat{a}^{\dagger}$ and $\hat{b}^{\dagger}$ for the modes $\ket{a}$ and $\ket{b}$ looks as follows 
\begin{equation*}
\hat{a}^{\dagger}\hat{b}^{\dagger}\ket{00}\stackrel{\hat{B}\hat{\varphi}\hat{B}}{\longrightarrow} 
\frac{1}{4}\left[-\textrm{i}(e^{-2\textrm{i}\varphi}-1)\hat{e}^{\dagger}\hat{e}^{\dagger}
+i(e^{-2\textrm{i}\varphi}-1)\,\hat{f}^{\dagger}\hat{f}^{\dagger}
-2(e^{-2\textrm{i}\varphi}+1)\,\hat{e}^{\dagger}\hat{f}^{\dagger}\right]\ket{00}\,.
\end{equation*}
For completely distinguishable photons (indicated by `$\sim$') the following relation applies
\begin{equation*}
\hat{a}^{\dagger}\tilde{b}^{\dagger}\ket{00}\stackrel{\hat{B}\hat{\varphi}\hat{B}}{\longrightarrow} 
\frac{1}{4}\left[-\textrm{i}(e^{-2\textrm{i}\varphi}-1)\hat{e}^{\dagger}\tilde{e}^{\dagger}
+i(e^{-2\textrm{i}\varphi}-1)\,\hat{f}^{\dagger}\tilde{f}^{\dagger}
-(e^{-\textrm{i}\varphi}-1)^2\,\hat{e}^{\dagger}\tilde{f}^{\dagger}
-(e^{-\textrm{i}\varphi}+1)^2\,\hat{f}^{\dagger}\tilde{e}^{\dagger}\right]\ket{00}\,.
\end{equation*}
The phase dependent coincidence probability is then given by the sum of the individual cases ($\mathcal{P}^{\textrm{}}_{\textrm{D1},\textrm{D2}}$ for identical, $\tilde{\mathcal{P}}^{\textrm{}}_{\textrm{D1},\textrm{D2}}$ for distinguishable photons), weighted with the indistinguishability visibility $V_{\textrm{HOM}}$:
\begin{align*}
\mathcal{P}^{\textrm{exp}'}_{\textrm{D1},\textrm{D2}}=
\underbrace{\frac{V_{\textrm{HOM}}}{2}(1+\cos(2\varphi))}_{\mathcal{P}^{\textrm{}}_{\textrm{D1},\textrm{D2}}}
+\underbrace{\frac{1-V_{\textrm{HOM}}}{4}(3+\cos(2\varphi))}_{\tilde{\mathcal{P}}^{\textrm{}}_{\textrm{D1},\textrm{D2}}}
=\frac{1}{4}\left[(3-V_{\textrm{HOM}})+(1+V_{\textrm{HOM}})\cos(2\varphi)\right]\,.
\end{align*}
For a perfect beam overlap, the reduction of the N00N state visibility strongly depends on the two-photon interference visibility $V_{\textrm{HOM}}$ and is given by $V_{N=2}=(1+V_{\textrm{HOM}}) / (3-V_{\textrm{HOM}})$.
Spatial modes which do not overlap perfectly are treated as a beam splitter with a $(1-\eta)$\,:\,$\eta$ splitting ratio\,\cite{Kacprowicz2010}.
Here, the reflected modes $\ket{\alpha}$, $\ket{\beta}$ of the operation $\hat{\eta}'$ (gray lines) and $\ket{\gamma}$, $\ket{\gamma_{1,2}}'$, $\ket{\delta}$, $\ket{\delta_{1,2}}'$ of the operation $\hat{\eta}''$ (purple lines) are feed back into the next BS in the photon path as non-overlapping modes, experiencing only classical correlations.
Including this additional non-overlapping contributions, the calculation of the coincidence correlation probability on the detectors D1 and D2 at $\tau=0$ leads to the phase dependent relation    
\begin{equation*}
\mathcal{P}^{\textrm{exp}}_{\textrm{D1},\textrm{D2}}=\frac{1}{4}\left(2+\eta^2(1-\eta^2V_{\textrm{HOM}})+\eta^2(1+\eta^2V_{\textrm{HOM}})\cos(2\varphi)\right)\,,
\label{eq:N_2N_2_exp}
\end{equation*}
already presented in the main paper.
$\eta=\sqrt{\eta'\eta''}$ combines the two individual overlaps on BS$'$ and BS$''$, $V_{\textrm{HOM}}$ defines the fraction of utilizable indistinguishable photons.

\begin{figure}[H]
\begin{center} 
\includegraphics[width=\linewidth]{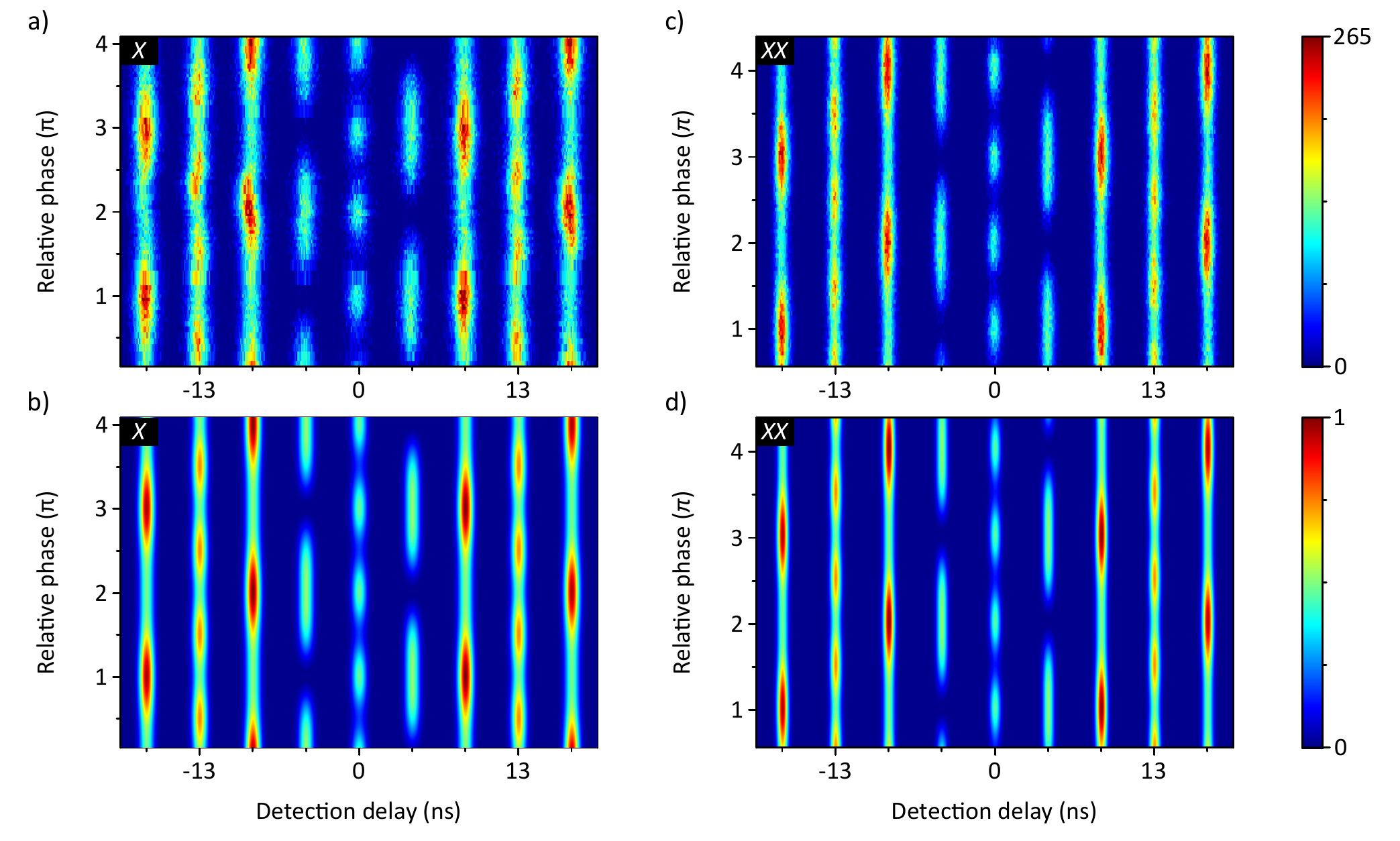}
\caption{
Complete phase dependence of the measured \textit{X} (a)) and \textit{XX} (c)) coincidence histograms. 
At zero delay, the predicted frequency doubling indicates the successful generation of the biphotonic N00N-state. 
The simulations (\textit{X} in b), \textit{XX} in d)) are in very good agreement with the experimental results and can reproduce the phase dependency for all detection delays.
\label{smfig3}} 
\end{center} 
\end{figure}

\subparagraph{N00N-state interference pattern}
As already mentioned in the main paper, the two consecutive photons (\textit{X} or \textit{XX}) arrive with a time separation of $\Delta t=4.4$\,ns at BS1.
Because of the 13.1\,ns repetition period, coincidence events from photons between different excitation cycles overlap in the histogram. 
In a theoretical consideration we can remove these additional coincidences and focus on the simple five peak pattern, originating from the statistics obtained by sending a double pulse into an unbalanced Mach-Zehnder interferometer.
Fig.\,\ref{smfig2}\,b),c) show the simulated phase dependent coincidence histograms.
For a phase of $\varphi=\pi/2$ the output probabilities of BS2$''$ are identical to the ones obtained after BS2$'$, therefore revealing a typical Hong-Ou-Mandel type histogram (Fig.\,\ref{smfig2}\,b)).
By turning the phase plate the coincidences for all time differences start to oscillate.
Besides the double frequently oscillating middle peak for a N00N-state, also the peaks for the time difference of $\pm\Delta t$ reveal an oscillating behavior with twice the single photon frequency (Fig.\,\ref{smfig2}\,c)).
However, these two photons do not form a N00N-state but travel one by one through the Sagnac interferometer.  
Referring to the experimental results, these double frequently occurring $\pm\Delta t$ coincidences are hidden due to the above mentioned overlap of different excitation cycles.
Fig.\,\ref{smfig3} shows a pseudocolor plot of the simulated (top) and measured (bottom) histograms for \textit{X} (left) and \textit{XX} (right) photons, including these coincidence overlaps.
The data is very well reproduced by the simulations, and also allows for a normalization as shown in Fig.\:\!3 of the main paper.

\newpage

